\documentclass[twocolumn,showpacs,preprintnumbers,floatfix,pra,aps]{revtex4-1}

\usepackage{graphicx}
\usepackage{dcolumn}
\usepackage{amsmath}
\usepackage{bbm}
\usepackage{hyperref}

\begin{document}
\title{Directed motion of doublons and holes in periodically driven Mott insulators
}
\author{Maximilian Genske$^1$}
\email{genske@thp.uni-koeln.de}
\author{Achim Rosch$^1$}
\affiliation{$^1$ Institut f\"ur Theoretische Physik, Universit\"at zu K\"oln, D-50937 Cologne, Germany\\
}
\date{\today}

\begin{abstract}
Periodically driven systems can lead to a directed motion of particles. We investigate this ratchet effect for
a bosonic Mott insulator where both a staggered hopping and a staggered local potential vary periodically  in time. If driving frequencies are  smaller than the
interaction strength and the density of excitations is small, one obtains effectively a one-particle quantum ratchet describing the motion of doubly occupied sites (doublons) and empty sites (holes). Such a simple quantum machine can be used to manipulate the excitations of the Mott insulator.
 For suitably chosen parameters, for example, 
holes and doublons move in opposite direction. To investigate whether the periodic driving can be used to 
move particles ``uphill'', i.e., against an external force, we study the influence of a linear potential $- g x$. 
For long times, transport is only possible when the driving frequency $\omega$ and the external force $g$ are commensurate, $n_0 g = m_0 \omega$, with $\frac{n_0}{2},m_0 \in \mathbbm{Z}$.
\end{abstract}

\pacs{67.85.-d, 05.60.Gg}
\maketitle

\section{Introduction}
One of the most simple quantum machines is a Hamiltonian where parameters are changed periodically in time.
If the parameters are changed slowly, one can realize a quantized adiabatic pump \cite{Thouless83}, where in each cycle all particles of an insulator are moved by one lattice constant. 
More recently, it has been shown that one can use the theory of topological insulators to classify such adiabatic pumps \cite{Hasan2010}.

More generally, the concept of a ratchet describes that in a periodically driven system, one can obtain directed motion even if all forces vanish on average \cite{Reimann2002,Haenggi2009}.
The necessary condition for this effect to emerge is the absence of inversion and time-reversal symmetries.
Generally, the ratchet mechanism can be used to explain a broad range of phenomena including not only man-made machines but also, for example, biological motors \cite{Julicher1997}.
In solid state systems ratchet effects have -- among others -- been used to experimentally realize an electron-pump \cite{Linke1999}, to create spin-currents \cite{Costache2010} or to detect asymmetries in graphene \cite{Drexler2013}.
Due to their high level of controllability, systems made of cold atoms in optical lattices have been an ideal test bed for implementing and investigating ratchets.
The very first of these (cold atom) ratchets relied on dissipative processes breaking the time-inversion symmetry \cite{Schiavoni2003,Gommers2005}.
The rapid experimental progress in the field of quantum optics allows now, however, to realize quantum-coherent ratchets by
manipulating atoms in time-dependent optical lattices created by the standing waves of lasers. 
A recent experiment \cite{Salger2009} demonstrated this effect by loading an atomic Bose Einstein condensate into a periodically changing potential of the form
$V_0 \cos(k x) \cos(\omega t)+V_1 \cos( 2 k x + \phi_1) \cos(\omega t+\phi_2)$. 
As soon as both $\phi_1$ and $ \phi_2$ deviate from  $0$ and $\pi$, such that inversion and time-reversal symmetries are broken, the atomic cloud drifts with an average velocity. 
Such a Hamiltonian quantum ratchet \cite{Schanz2005,Denisov2007,Ponomarev2009} works without any friction. 
This has to be generally contrasted with (semi-) classical ratchets, which mostly build on dissipation \cite{Reimann2002,Haenggi2009}.

A main motivation to study ratchets is to find new ways to manipulate quantum systems, but periodically driven Hamiltonian systems can also be used to probe properties of the system. 
The response of cold-atom systems to periodically driven systems has been intensively studied both theoretically \cite{Kollath2006, Cazalilla2011} and experimentally \cite{Stenger99,Stoferle04,Clement09}. 
For example, with the  so-called Bragg spectroscopy, one can measure the excitations of a system by studying Hamiltonians characterized by a simple time dependence, $H(t)=H_0+ H_1 \cos(\omega t)$.
In this case, however, no steady state currents can occur (due to an effective time-reversal invariance, $t \to -t$). 

The standard approach to describe periodically driven systems is the Floquet theory \cite{Shirley, Sambe}.
The Floquet theory \cite{Shirley, Sambe}, introduced in more detail below, uses that in a system where the Hamiltonian has a discreet translational invariance in time, $H(t)=H(t+T)$, the energy is still conserved modulo $\omega=2 \pi/T$.
In complete analogy to the band-theory of solids, one can therefore write the solution to the Hamiltonian as $\psi(t)=e^{-i \varepsilon t} \psi_{F}(t)$, 
i.e., as a product of a (temporal) plane wave and a time periodic function, $\psi_{F}(t+T)=\psi_{F}(t)$.
As stated above, for such a periodic system, energy is still conserved modulo $\omega$. 
The remaining conservation of energy does, however, strongly constrain the dynamics of the quantum system, especially in the presence of external forces.

In this paper, we show how single excitations of a bosonic Mott insulator, either doubly occupied sites (doublons) or empty sites (holes), can be manipulated using effectively one-particle quantum ratchets.
After the introduction of the model and a short review of the Floquet approach, we  present our results for the different dynamics of doublons and holes obtained from periodically driving the (staggered) lattice potentials.
When all necessary symmetries are broken, asymmetric Floquet states, $\psi_{F}$, support unidirectional motion of doublons and holes which can be independently controlled.
The remaining part of the paper investigates the effect of an additional static force, $g$.
We investigate under which condition the ratchet effect persists and discuss the importance of the
commensurability condition $n_0 g = m_0 \omega$, with $\frac{n_0}{2},m_0 \in \mathbbm Z$.
Such a condition is, e.g., also needed to obtain dynamical localization in kicked rotor systems \cite{Ringot2000} or to guarantee ballistic transport in electric quantum walks \cite{Genske2013}.
Furthermore, we show that the direction of transport can also be controlled in the presence of $g$, both  ``uphill'' and ``downhill'' motion can be achieved.

\section{Model and Floquet description}

\begin{figure}[t]
    \centering
    \includegraphics[width=7.5cm]{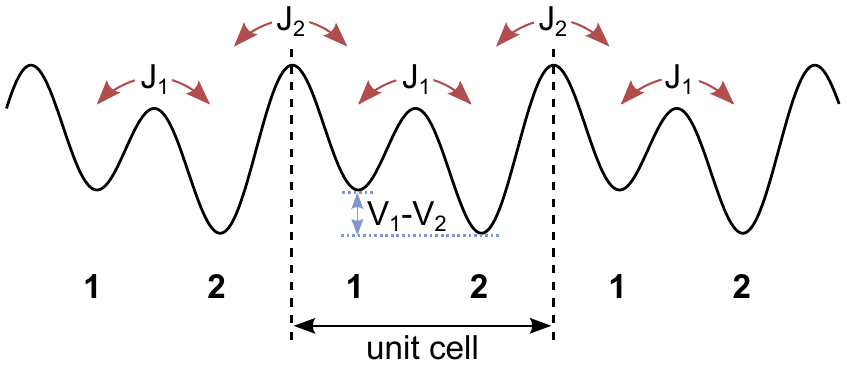}
    \caption{Illustration of the lattice model at a fixed time $t$. 
    Both the local hopping parameter and potential is staggered.
    $J_1,J_2,V_1,V_2$ are defined according to Eqs.~(\ref{J}) and (\ref{V}).
    }
    \label{fig1}
\end{figure}

We consider a Bose-Hubbard Hamiltonian with time- and space-dependent hopping strengths and external potentials of the form
\begin{equation}
\label{Ham1}
\begin{split}
 H(t) =- \sum_{ i}  J_{i}
 (t) (&a^{\dagger}_{i+1}a_{i}+h.c.) \\
 &+ \frac{U}{2} \sum_i n_{i} (n_{i} - 1) 
 + V_{i}
 (t) n_{i}
 \end{split}
\end{equation}
where $a_{i}^{\dagger}$($a_{i}$) creates (annihilates) a particle at site $i$, $n_i$ is the particle number operator, 
$J_{i}$ is a hopping matrix element, $V_i$ is an external potential and $U$ describes the on-site interaction strength. 
We use units where $\hbar=1$ and the lattice spacing is set to $1$ throughout the paper.
As in the experiments of Salger {\it et al.} \cite{Salger2009}, we consider a case where the effective optical lattice is described by two wave-lengths $\lambda$ and $2 \lambda$ such 
that a staggered hopping and staggered potential are created, which are both modulated by the same frequency.
Adding also the effects of a uniform force $g$ (see below) we consider
\begin{align}
\label{J}
J_i(t)&=\left\{ 
\begin{array}{ll}
J_{s,o}+\cos(\omega t) J_o  & \text{for}\ i\ \text{odd} \\
J_{s,e}+\cos(\omega t)  J_{e} & \text{for}\ i\ \text{even} 
\end{array}
\right. \\
\label{V}
V_i(t)&=\left\{ 
\begin{array}{ll}
-g i + V_s + \cos(\omega t+\varphi ) V_o  & \text{for}\ i\ \text{odd} \\
-g i - V_s + \cos(\omega t+\varphi)  V_{e} & \text{for}\ i\ \text{even} 
\end{array}
\right.
\end{align}
A sketch of the effective hoppings and potentials is given for $g=0$  in Fig.~\ref{fig1}.
Note that there is a relative phase shift $\varphi$ in the time dependence of the oscillating hopping terms ($J_e$, $J_o$) 
and the corresponding potential terms ($V_e$, $V_o$). 
Above, we have written the 1d version of the problem. 
All of the results presented below apply equally also for dimensions $d>1$ as long as the modulations of $H$ occur only in one direction of space.

We consider the limit where the onsite-interaction $U$ is considerably larger than the hopping rates,
the frequency $\omega$ and the initial temperature of the system (before the ratchet is switched on), $U\gg J_i, V_i, T$. 
Furthermore, we consider an average occupation of the system by one boson per site.  
In this limit the main excitations of the system are doubly occupied sites (doublons) and empty sites (holes) with creation operators $d_i^\dagger$ and $h_i^\dagger$, respectively. 
We assume that they are so diluted that we can neglect their scattering on the time scale of the experiment.
Therefore their single-particle dynamics is described by non-interacting Hamiltonians, which 
to leading order in $1/U$ have the simple form
\begin{align}
 H_{d}(t) &= \sum_{i} - 2 J_{i}(t) (d^{\dagger}_{i+1} d_{i} + d^{\dagger}_{i}d_{i+1}) + V_{i}(t) n_{di} \label{Hd}  \\
 H_{h}(t) &=\sum_{i} - J_{i}(t) (h^{\dagger}_{i+1}h_{i} + h^{\dagger}_{i} h_{i+1}) - V_{i}(t) n_{hi} \label{Hh}
\end{align}
In two aspects the dynamics of doublons and holes is different: First, the hopping rate of doublons is twice the hopping rate of holes (due to stimulated hopping arising from Bose statistics), second, the external potential is reversed as a hole is a missing boson.

In the absence of external forces, $g=0$, the system can be described by a two-site unit cell and therefore two bands in $k$ space, $-\frac{\pi}{2} \leq k < \frac{\pi}{2}$. Defining 
$d^\dagger_{n,e}=d^\dagger_{2n}$, $d^\dagger_{n,o}=d^\dagger_{2n+1}$ and the Fourier transformation 
$d^\dagger_{k,e/o}=\sum_n d^\dagger_{n,e/o} e^{i k 2 n}$, the Hamiltonian is conveniently described
by simple $2 \times 2$ matrices $H_k^n$ oscillating with $e^{i \omega n t}$ 
\begin{equation}
 H_d(t)  =\int \frac{dk}{2 \pi}\, (d^\dagger_{k,o}, d^\dagger_{k,e}) \left(\sum_n H_{d,k}^n e^{i n \omega t} \right) \begin{pmatrix}d_{k,o} \\ d_{k,e}  \end{pmatrix}\label{momentumB}
\end{equation}
with
\begin{align}
\label{H0}
H_{d,k}^0&= \begin{pmatrix}
 V_s & -2 (J_{s,o}+ J_{s,e}  e^{- 2 i k}) \\
-2 (J_{s,o}+ J_{s,e} e^{2 i k})& -V_s 
 \end{pmatrix}
\\
\label{H1}
H_{d,k}^{\pm 1}&= \frac{1}{2} \begin{pmatrix}
 e^{\pm i\varphi} V_o & -2 (J_o + J_e  e^{-2 i k}) \\
-2 (J_o + J_e e^{2 i k}) &e^{\pm i\varphi} V_e
 \end{pmatrix}
\end{align}
Similar equations hold for the dynamics of holes.

Generally, time dependent problems are not readily solvable.
However, if the system is periodic in time with periodicity $T=\frac{2 \pi}{\omega}$, the Floquet theorem can be applied: $|\psi_{\alpha}(t)\rangle = e^{-i\varepsilon_{\alpha} t} |\psi_{F,\alpha}(t)\rangle$, 
where $|\psi_{F,\alpha}(t)\rangle$ is a Floquet state with   $|\psi_{F,\alpha}(t)\rangle= |\psi_{F,\alpha}(t+T)\rangle$ and $\varepsilon_{\alpha} \in [-\omega/2,\omega/2)$ is its corresponding quasienergy. 
The index $\alpha$ refers to the initial Hilbert space structure.
Introducing the Fourier components
$|\psi_{F,\alpha m}\rangle= \int_{t_0}^{t_0+T} dt |\psi_{F,\alpha}(t) \rangle e^{-i \omega m t}$, one obtains an effective time-independent Schr\"odinger equation in Floquet space of the form~\cite{Shirley, Sambe}
\begin{equation}
\varepsilon_{\alpha}  |\psi_{F,\alpha m}\rangle = \sum_{m'} \big( H^F_{d/h} \big)^{m m'} |\psi_{F,\alpha m'}\rangle
\end{equation}
with the Floquet Hamiltonian 
\begin{equation}
\begin{split}
 \big( H^F_{d/h} \big)^{m m'} = &~m~\omega~\delta_{m m'}\mathbbm{1} \\ &+ \frac{1}{T} \int_{t_0}^{t_0+T} dt~e^{-i (m-m')\omega t} H_{d/h}(t) 
\end{split}
\label{eq:Floquet}
\end{equation}
Note that $H^F$, and hence the dynamics of the system, is generally dependent on the initial time $t_0$.
For the rest of the paper we set $t_0 = 0$, since conclusions are not affected qualitatively.

Effectively, we have enlarged the Hilbert space by introducing the extra Floquet index $m$. 
In the basis discussed in Eq.~(\ref{momentumB}), we obtain $(H^F_d)^{m+n,m}=H_{d,k}^n$: the oscillating terms provide an effective hopping in Floquet space.
Exact diagonalisation allows to obtain the precise form of Floquet eigenstates and quasienergies, and, hence, information about the system at arbitrary times.
For practical applications the Floquet Hamiltonian needs to be truncated.
The first term in Eq.~(\ref{eq:Floquet}) linear in $\omega$ ensures that modes with very large $m$ are suppressed and therefore it is
sufficient to include only a finite number $M$ of modes
(larger than $D/\omega$ where $D$ is the maximal energy scale in the problem) to obtain numerically exact results. It is also sufficient to obtain
only the eigenvectors with eigenvalues in the interval $[-\omega/2,\omega/2)$ in the middle of the spectrum of  $H^F_{d/h}$, which are least affected by the truncation, using that all other eigenvectors and eigenenergies of the Floquet Hamilitonian (\ref{eq:Floquet}) can be obtained by a simple translation in Floquet space and an increase of the energy by $\omega$.

The time evolution operator $U(T)$ for the time interval $T$ can be written as
\begin{equation}
U(T)=\frac{1}{M} \sum_{m,m'} \left( e^{-i H^F_{d/h} T}\right)_{m,m'}=e^{-i H^{\rm eff}_{d/h} T}
\label{heffF}
\end{equation}
where $H^{\rm eff}_{d/h}$ is an effective time-independent Hamiltonian describing the stroboscopic time evolution \cite{Goldman2014,Kitagawa10} associated with the time interval $T$.
Such effective Hamiltonians can be used to investigate a broad range of interesting phenomena such as artificial gauge fields (see below) or topological states of matter \cite{Kitagawa10,Lindner2011,Grushin2014}.

\section{Ratchet effects in the uniform system}

\begin{figure}[t!]
    \centering
    \includegraphics[width=8.5cm]{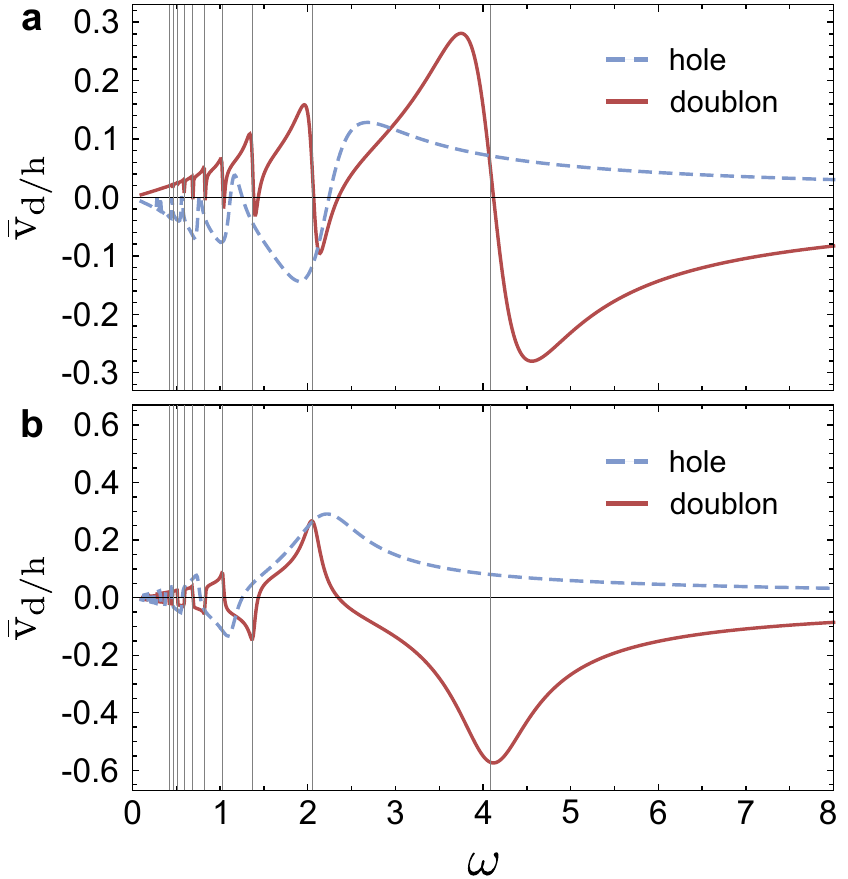}
    \caption{Doublon (red/solid) and hole (blue/dashed) asymptotic velocity for (a)~the sudden switching and (b)~the adiabatic case.    
    Parameters: $\varphi=-\pi/2$, $J_{s,e}=0.67$, $J_{s,o}=0.33$, $J_{e}=0.67$, $J_{o}=-0.17$, $V_{s}=0.4$, $V_{e}=0.4$ and $V_{o}=-0.4$.
    Vertical lines indicate the first ten doublon resonances, i.e., $\omega=\Delta E/n$, $n \in \{1,...,10\}$.}
    \label{fig:fig2}
\end{figure}

The central objective of this paper is to quantify the ratchet effect.
This is done by determining the velocity of the doublons and holes 
\begin{align}
\label{veld}
 v_d(t)&= i \frac{1}{N_d} \sum_{n} 2 J_{n}(t) [d^{\dagger}_{n+1}d_{n} - d^{\dagger}_{n}d_{n+1}] \\
 \label{velh}
 v_h(t)&= i \frac{1}{N_h} \sum_{n} J_{n}(t) [h^{\dagger}_{n+1}h_{n} - h^{\dagger}_{n}h_{n+1}] 
\end{align}
obtained by dividing the current by the density ($N_d=\sum_i d^\dagger_i d_i$ and $N_h=\sum_i h^\dagger_i h_i$ are the total number of doublons and holes, respectively). 
Note that we use conventions where a positive $v_{d/h}$ describes doublons or holes moving to the right. 
Transforming operators \ref{veld},\ref{velh} to Floquet space, $v_{d/h}(t)\to v_{d/h}^F$, allows to calculate the velocity by taking the expectation value with respect to the system expressed in terms of Floquet states. 
The asymptotic velocity, i.e., the average velocity in the limit $t \to \infty$, can be obtained by considering only
diagonal contributions of a single Floquet mode, i.e., $\bar{v}_{d/h} = \sum_{\alpha} p_\alpha \langle \psi_{F,\alpha} | v_{d/h}^F | \psi_{F,\alpha} \rangle$, as interference terms of two Floquet states with different energy average to zero.
Here, $p_\alpha$ is the probability that the Floquet band $\alpha$ is occupied (normalised to 1).
Alternatively, the asymptotic velocity can be determined by the slope of the quasienergy bands, i.e., $\bar{v}_{d/h}(k)=\sum_\alpha p_\alpha  \partial \varepsilon_{\alpha}(k) / \partial k$. 
This is fully analogous to the conventional semiclassical picture of transport associated with energy bands, 
and can be analytically shown by applying the Hellmann-Feynman Theorem to the Floquet theory \cite{Schanz2005}.
Breaking the inversion symmetries of the system simultaneously breaks the symmetry of the quasienergy bands, thus, allowing for a non-zero velocity for initial states with momentum $k=0$.
In fact, specific band shapes can be tailored by changing the parameters appropriately.

The Floquet bands are still periodic in momentum space and therefore $\epsilon_\alpha(-\pi)=\epsilon_\alpha(\pi)+ n_\alpha\,  \omega$ with $n_\alpha \in \mathbbm Z$. 
While the goal of the paper is to investigate ratchets for dilute doublon and hole gases, we briefly review the consequences of our discussion for insulating states, i.e., 
Mott insulators or fermionic band insulators, where one can realize topological quantum pumps.  
In this case one has to investigate the current (or particle velocity) when all states of a given band are occupied. 
The average velocity of band $\alpha$ is given by  $\int \frac{dk}{2 \pi} \partial \varepsilon_{\alpha}(k) / \partial k=n_\alpha \, \frac{\omega}{2 \pi}=\frac{n_\alpha}{T}$ and therefore quantized:  
within each cycle of the ratchet each particle moves  $n_\alpha$ lattice spacings to the right \cite{Thouless83}.
Such a pump is, however, only stable in the adiabatic limit \cite{Thouless83, Kitagawa10}, but not for large frequencies. 
As has been shown in Refs.~\cite{Kitagawa10, Qi10}, $n_\alpha$ can be identified with a topological winding number. 
The sum of winding numbers over all bands vanishes for all models with local hopping parameters, $\sum_\alpha n_\alpha=0$.
Therefore it is unavoidable, that the energies (modulo $\omega$) of a band with 
$\epsilon_{\alpha}(-\pi)=\epsilon_{\alpha}(\pi)+  n_\alpha \omega$ and $n_\alpha>0$ will cross some other band
with $\epsilon_{\alpha'}(-\pi)=\epsilon_{\alpha'}(\pi)+n_{\alpha'} \omega$ with $n_{\alpha'}<0$. 
A small perturbation is therefore sufficient to hybridize the two bands, thereby changing their winding number. 
Such a band crossing can, however, be exponentially suppressed in the adiabatic limit, i.e., 
when $\omega$ is much smaller than the relevant band gap.
See e.g. Ref.~\cite{Kitagawa10} for a comprehensive discussion of topological Floquet bands.

We proceed by investigating the case where initially the temperature is low compared both to the bandwidth and the Mott gap. 
Therefore only a few doublons and holes are present, occupying states with momenta close to zero. 
We assume that for $t<0$ $V_s=V_e=V_0=J_e=J_o=0$ and study the evolution of the $k=0$ state.
At $t=0$ both the static staggered potential $V_s$ and the oscillating terms are switched on.
Values of parameters used throughout this paper can be found in caption of Fig.~\ref{fig:fig2}.
We consider two limits: in the first case we consider a sudden switching on of the oscillating terms, $H=H_0+\Theta(t) (H(t)-H_0)$,
in the second case an adiabatic switching on, $H=H_0+f(t) (H(t)-H_0)$ where $f(t)$ smoothly changes from 0 to 1 on a time scale assumed to be much larger than all other time scales of the problem. 
In the latter case, the system evolves smoothly into a Floquet eigenstate while in the first one a superposition of those has to be considered.

Finite net velocities are generically expected whenever they are allowed by symmetry.
Due to the presence of {\em both} staggered potential and staggered hopping all mirror symmetries are broken and for $\varphi \neq 0,\pi$ there exist also no time-reversal symmetry.
In Fig.~\ref{fig:fig2}(a) we show the velocity as function of frequency after a sudden switching on 
of large oscillating terms. For both doublons and holes the response is characterized by 
a number of resonances at frequencies $\omega \approx \Delta E/n$, with $\Delta E$ corresponding to the $k=0$ energy difference between two bands of the (undriven) system.  
While for $n=1$ there is only one sign change, there are always two sign changes associated with  $n\geq2$. 
These sign changes persist even in the limit where the oscillating terms in the Hamiltonian are very weak.
These effects can be easily understood from perturbation theory which is shown for $n=1$ in appendix~\ref{appPT}.
For $\omega \approx \Delta E/n$ two entries of the Floquet matrix become degenerate leading
in perturbation theory to a pole in the response of the velocity which is smeared out when in degenerate perturbation theory the effect of a finite off-diagonal matrix element (i.e., a finite oscillating term in the Hamiltonian) is considered. 
If these oscillating terms are controlled by a small parameter $\epsilon$, the height of the observed peaks are $\mathcal O(\epsilon^n)$ (see appendix~\ref{appDPT}).
Since the poles have the same sign for all $n$, the observed pattern of sign changes emerges.

\begin{figure}[t!]
    \centering
    \includegraphics{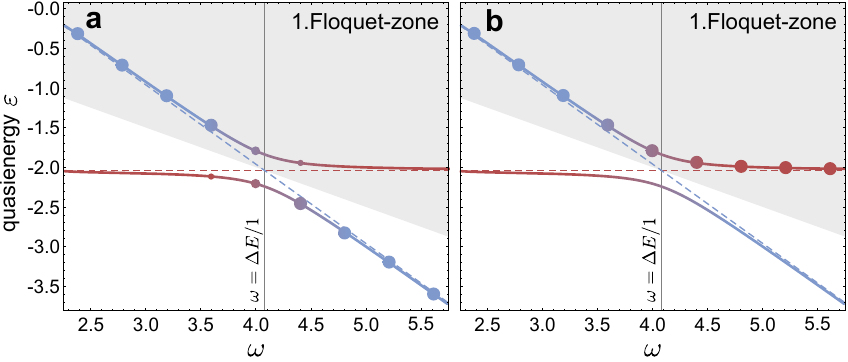}
    \caption{Avoided crossing at the main resonance $(\omega=\Delta E/1)$ of the doublon bands (parameters as in Fig.~\ref{fig:fig2}) at the lower edge of the first Floquet-zone (gray shaded area). 
    Dashed lines correspond to eigenenergies of $H^{F}$ in the absence of time-dependent terms in the original Hamiltonian.
    Points illustrate population of the respective band for (a)~the diabatic and (b)~the adiabatic case.
    In (a)~the system changes the band as function of $\omega$, while in (b)~it remains in the Floquet band.}
    \label{fig:avcross}
\end{figure}

Remarkably, the frequency dependence for an adiabatic switching on of the net velocity takes a completely different form, see Fig.~\ref{fig:fig2}(b).
Here, for $n=1$ there is no sign change at all, but there is always one sign change for $n \geq 2$.
This behaviour reflects the avoided level crossing occurring at each resonance: 
when changing from $\omega< \Delta E/n$ to $\omega > \Delta E/n$ one effectively changes the band for a sudden switching on of the ratchet while one stays in the same band in the adiabatic case, see Fig. \ref{fig:avcross}.
We do not consider the case of a finite ramping speed of the oscillating potential which maps to the classical Landau-Zener tunneling problem in the limit when the ramping up time is large compared to the oscillation period $T$.

For large frequencies one can use that the first term in the Floquet Hamiltonian (\ref{eq:Floquet}) becomes large, suppressing all matrix elements connecting different Floquet blocks. 
Therefore perturbation theory (shown in appendix~\ref{appPT}) in the oscillating terms becomes valid for $\omega\gg J_o,J_e,V_o, V_e$ even for large  $J_o,J_e,V_o, V_e$. 
The velocity in this limit decays with $1/\omega$. One can also interpret this result as arising from an effective vector potential in the effective Hamiltonian defined in Eq.~(\ref{heffF}). 
A vector potential shifts the momentum, giving rise to a finite velocity at $k=0$, i.e., $\varepsilon_{\alpha}^0(k)\to\varepsilon_{\alpha}^0(k-\mathcal A_d)$, 
where $\varepsilon_{\alpha}^0$ describes the eigenenergies of the undriven system.
The effective (doublon) vector potential can be explicitly determined in the present case to read (see appendix \ref{appEVP})
\begin{equation}
 \mathcal A_d = \frac{1}{2 \omega} \sin(\varphi) (V_o - V_e) \left( \frac{J_o}{J_{s,o}} - \frac{J_e}{J_{s,e}} \right)
\end{equation}
Since the potential terms change sign when replacing doublons with holes, it is clear that also this effective vector potential has an opposite sign for both species.
Such effective vector potentials have recently been intensively studied both experimentally \cite{Lin2009,Lin2011,Struck2012,Beeler2013,Aidelsburger2013,Miyake2013} and theoretically \cite{Sorensen2005,Hauke2012,Celi2014}.

As doublons and holes have different hopping rates, also the relevant resonance frequencies are different.
Thus by tuning the frequency of the oscillating potential, one can easily reach situations where, e.g., holes are driven to the right but doublons move in the opposite direction. 
Similarly, one can also reach situations where only one of the two species is moving. 
We have therefore shown, that with the help of ratchets one can selectively control the various excitations of a homogeneous interacting many-particle system. 
Using focused lasers which change, e.g., the effective resonance frequencies of the ratchet locally, one can combine this with a spatial control of where the excitations are moving to.

\section{Ratchet effects in the presence of a force}

\begin{figure}[t!]
    \centering
    \includegraphics[width=7cm]{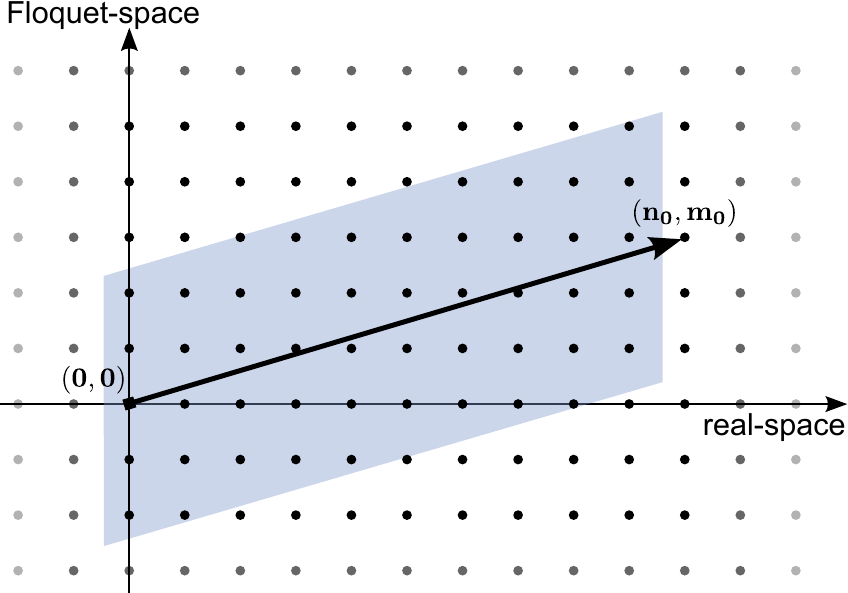}
    \caption{Illustration of states in a mixed real-space/Floquet-space representation.
    The vector from $(0,0)$ to $(n_0,m_0)$ fulfils the commensurability condition (\ref{eq:comm}).
    The practical truncation of the Floquet matrix is done around this vector and only states lying in the blue/shaded region are used in calculations.
    In most cases it is sufficient to take only a few Floquet states per real-space index into account.} 
    \label{fig:trunc}
\end{figure}

Can a ratchet also perform work and move the particles ``up the hill''? 
This is also an interesting question as in most cold atom realization an external parabolic potential holds the atomic cloud together.
To investigate such questions, we consider the effect of a constant force, $g \neq 0$ in Eq.~(\ref{V}).
The previous analysis cannot be carried over to this problem straightforwardly,
since the presence of the linear potential breaks the translational invariance of the original Hamiltonian.

The problem does, however, simplify considerably when the strength of the electric field and the driving frequency are commensurate with each other
\begin{equation}
  n_0 g = m_0 \omega \label{eq:comm}
\end{equation}
with integer $m_0 \in \mathbbm{Z}$ and even $n_0$, $\frac{n_0}{2} \in \mathbbm{Z}$ (due to the dimerized structure of our Hamiltonian). 
In this case, a shift of the particle by $n_0$ lattice sites to the right costs an energy which is a multiple of $\omega$. 
If we translate this property into Floquet language, this implies that there is a new `translation invariance' in mixed position/Floquet space: 
when moving $n_0$ steps to the right in real space and simultaneously $m_0$ steps in Floquet space, one recovers the same Hamiltonian. 
The eigenstates $|\Psi_k\rangle$ of the Floquet Hamiltonian can therefore be chosen as
\begin{equation}
\langle i,m|\Psi_k\rangle=e^{i k r_i} \langle i,m|u_k\rangle
\end{equation}
with $\langle i,m|u_k\rangle=\langle i+n_0,m+m_0|u_k\rangle$, where $i$ is the real-space and $m$ the Floquet index and 
$-\frac{\pi}{n_0}\le k < \frac{\pi}{n_0}$ the effective momentum (projected on the space coordinate for convenience). For each $k$ one obtains now $n_{0r}$ bands with energies $-\omega/2 \le \varepsilon_\alpha<\omega/2$,
where $n_{0r}$ is the reduced denominator of $m_0/n_0$.
As the strongest response arises from fractions with small denominators (see below), we combine for our plots data for $n_0=60$ with data for $n_0=14, 16$ and $18$ with arbitrary values of $m_0$ and $\omega$. 
This allows to cover all fractions with $\frac{n_0}{2}=1,2,...,10$. 
For an effective calculation especially for large forces it is important to choose a suitable truncation of the Floquet space which is best defined in a mixed real-space/Floquet-space representation shown in Fig.~(\ref{fig:trunc}): 
we consider only Floquet blocks close to the line connecting $(0,0)$ to $(n_0,m_0)$.

\begin{figure*}[t!]
    \centering
    \includegraphics[width=17.5cm]{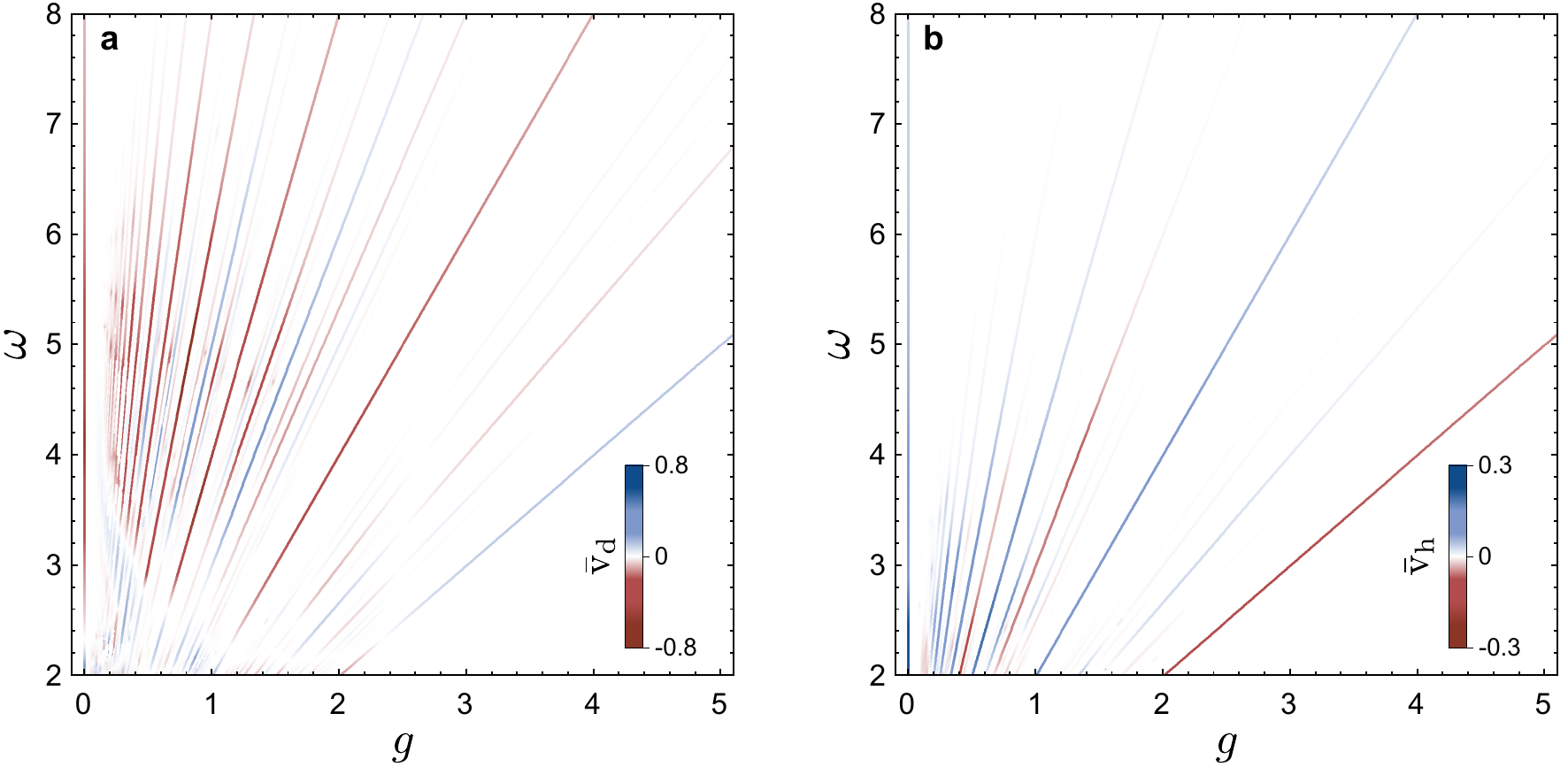}
    \caption{Asymptotic (a) doublon and (b) hole velocity as function of driving frequency $\omega$ and external force $g$ for an adiabatically switched on ratchet (parameters as in Fig.~\ref{fig:fig2}).
    Combined results are shown for real-space unit cell sizes $n_0 = 14,16,18$ and $60$ and arbitrary values of $m_0$.}
    \label{fig5}
\end{figure*}

Above, we have described the algorithm in a mixed real-space/Floquet-space picture. 
A fully equivalent formulation can also be obtained by a simple Gauge transformation which encodes the constant force (which can be viewed as an electric field) in a time-dependent vector potential. 
In this case, one replaces in the Hamiltonian~(\ref{Hd}) $d_{i+1}^\dagger d_i$ by $d_{i+1}^\dagger d_i e^{i g t}$ and in (\ref{Hh}) $h_{i+1}^\dagger h_i$ by $h_{i+1}^\dagger h_i e^{-i g t}$. 
This leads to a Hamiltonian which is translational invariant in real space but characterized by two frequencies. Therefore one obtains two instead of one Floquet indices.
Under the condition (\ref{eq:comm}) the two frequencies $\omega$ and $g$ are commensurate, allowing again for a simplified treatment equivalent to the one described above.

We consider an initial single particle excitation with zero momentum, $k=0$ (as above). At time zero both 
the electric field and the oscillating terms are switched on either instantaneously or adiabatically.
Figs. \ref{fig5}(a) and \ref{fig5}(b) show the velocity of doublons and holes in the long-time limit $t\to \infty$ calculated for commensurate $g=\frac{m_0}{n_0} \omega$ in the adiabatic case. 
Red colors describe left-moving and blue colors right-moving excitations. 
Depending on both frequency and electric field, the particles can move either `up' or `down'. 
By comparing the velocity of doublons and holes, 
one realizes that by choosing the appropriate values of $g$ and $\omega$ one can tune the parameters such that  doublons and holes move in the same or in opposite directions or that only one species is moving.

In general, the motion of the particles is most pronounced when $g/\omega=\frac{m_0}{n_0}$ can be written as a fraction with small integers $m_0$ and $n_0$.
To explain this observation, we first recall that for static band Hamiltonians a static external force does not lead to a finite net velocity of the particles: 
only Bloch oscillations can occur. 
This can be understood from energy conservation: as the kinetic energy is bounded, the system cannot absorb the potential energy it would gain in the presence of a finite net velocity. 
In the presence of interactions, this physics is discussed, e.g., in Ref.~\cite{Mandt2011}.
In the presence of the oscillating terms, the energy is still conserved modulo $\omega$. 
This implies that an energy conserving transport process is possible if the energy gain $-g n_0$ obtained by hopping $n_0$ steps to the right, is an integer multiple of $\omega$ as described by Eq.~\ref{eq:comm}. 
For large values of either $m_0$ or $n_0$, this process is, however, exponentially suppressed as the particle has to tunnel through long distances in either real- or Floquet space until the resonance condition is met. 
This implies that for $t \to \infty$ a finite average velocity is only possible if $g$ and $\omega$ are commensurate, while it vanishes for all incommensurate ratios (i.e., it is finite only on a set of measure zero). 

This situation is different when finite times $t$ are considered. In this case, we have computed the average velocity for a finite system. 
To be able to describe arbitrary (non-commensurate) values of $g$ while using the above described Floquet code, we implemented the following modified potential for a given site $n$
\begin{align}
V(g,n)=- g n - \sum_{i \in \mathbbm Z} \Theta(n-i n_0) \left(\frac{m(g) \omega}{n_0}-g\right) n_0
\end{align}
where the $\Theta$ function is defined by $\Theta(x)=1$ for $x \ge 0$ and $\Theta(x)=0$ for $x<0$. 
The integer $n_0$ describes the fixed size of an effective real-space unit cell (we use $n_0=120$) and $m(g)$ is an integer which is chosen to be the best approximation to $\frac{g n_0}{\omega}$.
With this construction, the potential has the commensuration property $V(g,n+n_0)=V(g,n)-m(g) \omega$ needed for our implementation. 
For values of $g$ which cannot be written in the form $g=\frac{m_0}{n_0} \omega$, the potential has jumps ($\Delta V< \omega/2$) at the boundaries of the unit cell, $n=i n_0$. Due to this finite size error, one cannot use this model to calculate the long-time asymptotic, but it is valid for short times if one determines
the velocity of the bosons only in the center of the unit cell, far away from the boundaries.
This guarantees that for short times, $t<n_0/(2 v)$, the result is not affected by these jumps (with exponential precision). 
Here $v$ is the Lieb-Robinson velocity \cite{Lieb1972}, the maximal velocity of signal propagation (of the order of twice the hopping rate). 
As an initial state we consider a doublon at zero momentum with wave function $|\psi(0)\rangle=1/N\sum_{n=1}^{N} | n \rangle$ in real space. At time $t=0$ the ratchet is suddenly switched on. 
To calculate  $|\psi(t)\rangle$, first $|\psi(0)\rangle$ is projected onto the Floquet eigenstates, then $|\psi(t)\rangle$ is evaluated using the Floquet theorem, 
and finally the expectation value of the velocity operator, $\langle \psi(t)| v_{d,n}(t) |\psi(t)\rangle$, is evaluated in the middle of the unit cell, $n=n_0/2$. 
Finally, we calculate the average velocity for the interval $[0,t]$,
\begin{equation}
\bar v(t)=\frac{N}{t} \int_0^t \langle \psi(t')| v_{d,n}(t') |\psi(t')\rangle dt'=\frac{\Delta R}{t} 
\end{equation}
Due to the averaging, oscillating contributions to $v$ are suppressed. 
For the chosen normalization $\bar v(t)$ directly gives the average velocity of the doublons in the interval  $[0,t]$ obtained from the shift $\Delta R$ of the center-of-mass of the wave function, $R=\sum_n n d^\dagger_n d_n/\sum_n d^\dagger_n d_n$ divided by $t$.

\begin{figure}[t!]
    \centering
    \includegraphics{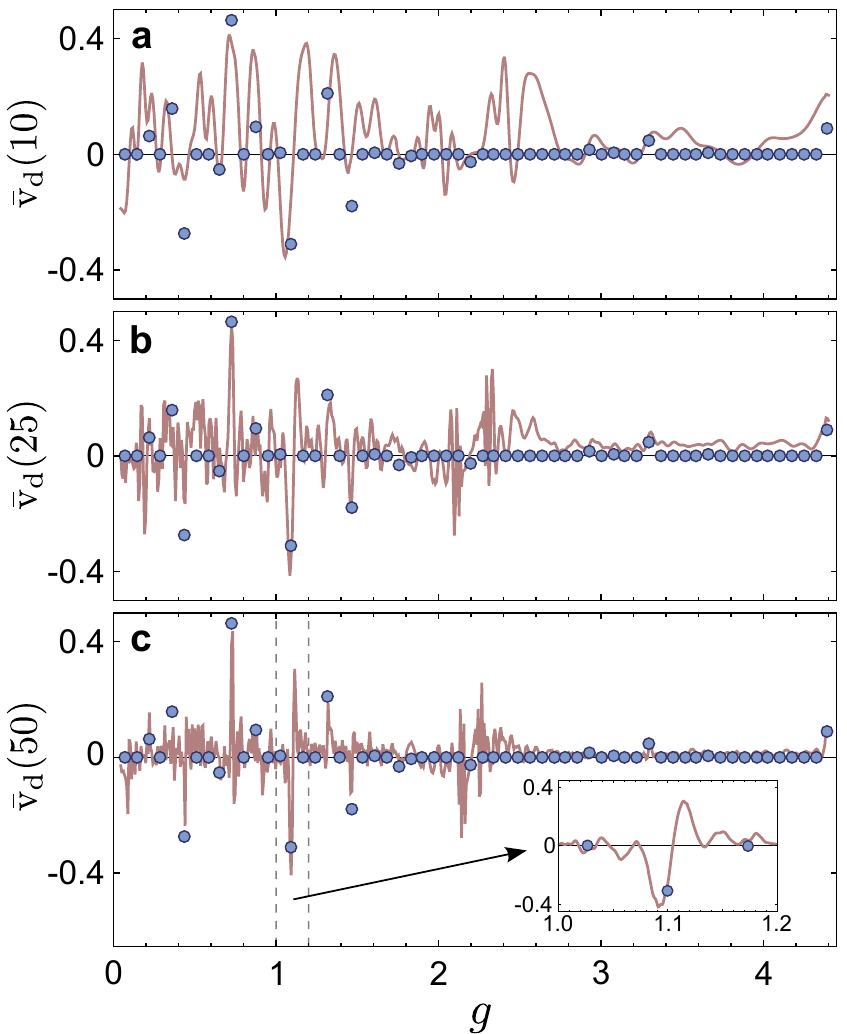}
    \caption{Average doublon velocity (solid lines) at $t=10, 25$ and $50$ after a sudden switching on of the ratchet potential as function of the external force $g$ for a fixed driving frequency $\omega = 4.4$ ($n_0=120$). 
    The discreet points at $g_c = \frac{m_0}{60}\omega$ show the average velocity in the limit $t \to \infty$.
    The inset in (c) reveals the characteristic pattern of minima and maxima caused by effective Bloch oscillations (see text) near the $\frac{m_{0r}}{n_{0r}}=\frac{1}{4}$ resonance.}
    \label{fig7}
\end{figure}

The average velocity as a function of the strength of the external force $g$ is shown in Fig.~\ref{fig7}.
The plot shows how for increasing time $t$ the average velocity (solid line) approaches the $t\to \infty$ limit (dots). The curves are characterized by oscillations with a width which shrinks with $1/t$. With shrinking width
only those electric fields give a sizable contribution which fulfil the commensuration condition (\ref{eq:comm}) with a precision determined by $1/t$,
$n_0 g = m_0 (\omega+ \mathcal O(1/t))$. Close to each point where the commensuration condition is fulfilled, there is a characteristic pattern of oscillations (see inset of Fig.~\ref{fig7}(c)).
These oscillations can be interpreted as Bloch oscillations of an effective Floquet bandstructure and of an effective electric field:
At the commensurate point, $g=g_c=\frac{m_0}{n_0} \omega = \frac{m_{0r}}{n_{0r}} \omega$, with $\frac{m_{0r}}{n_{0r}}$ being the reduced fraction, 
one obtains in the enlarged unit cell $n_{0r}$ Floquet bands, which include both the effect of the constant force $g$ and of the oscillating Hamiltonian. 
A finite $g-g_c$ acts like a small effective electric field for these Floquet bands. 
It induces Bloch oscillations whose period can be computed semiclassically using $\partial_t p=g-g_c$ and $T \partial_t p=\frac{2 \pi}{n_{0r}}$.
The period of these effective Bloch oscillations is therefore given by 
\begin{equation}
T=\frac{2 \pi}{n_{0r} |g-g_c|}
\end{equation}
As a function of $g$ one hence observes peaks in $\bar v(t)$ not only close to the commensurate points $g_c$, but there is a series of minima and maxima separated by
$\Delta g\approx  \frac{\pi}{n_0 t}$ as can be seen in the inset of Fig.~\ref{fig7}(c).

\section{Conclusion and Outlook}

In this paper, we have investigated periodically driven quantum systems focussing on simple lattice models
with and without a constant external force. As an example, we have studied a thermally activated Mott insulator 
characterized by a small number of doublon and hole excitations. By choosing the right frequency, one can selectively couple to either doublons and holes and move them across a lattice. More complicated is the situation in the presence of a constant external force. Within our lattice model, transport in the long-time limit is only possible if external drive and force are commensurate to each other.

For the future, it will be interesting to use tailored driven systems to realize all types of quantum machines. One natural goal could be to develop a cooling device which uses the selective transport of doublons and holes
to either remove them from the system or to bring them close to each other for their controlled recombination
by an appropriate time-dependent change of the Hamiltonian.

Our paper has completely neglected an important effect: the interaction of the excitations among each other.
While this approximation is valid for small densities of excitations, it breaks down when the density of excitation grows. 
A special situation arises in one dimension where doublons and holes generically cannot pass each other due to a strong local repulsion. 
Generically, the presence of interactions leads to heating in the long-time limit \cite{Eckstein2011}: 
ultimately, the only stationary state is the infinite temperature state, where all many particle states are occupied with equal probability. We expect, however,
that even in strongly interacting systems it will be possible to design quantum machines which on the one hand manipulate
a strongly interacting system efficiently and on the other hand do not produce too much entropy.

\begin{acknowledgments}
 This work is financially supported by the SFB TR 12 of the DFG, the Deutsche Telekom Stiftung (M.G.) and the Bonn-Cologne Graduate School of Physics and Astronomy (M.G.). 
\end{acknowledgments}

\appendix
\section{Perturbation theory}
\label{appPT}

We calculate the doublon velocity using a perturbative approach in the Floquet basis. Here, we express the full Floquet Hamiltonian $H^{F}$ as
\begin{equation}
\label{eq:HF}
H^{F} = H^{F}_0 + \epsilon_1 H^{F}_1 + \epsilon_2 H^{F}_2
\end{equation}
with $\epsilon_1,\epsilon_2 \ll 1$ and
\begin{align}
 H^{F}_{0,nm} &= (n\omega \mathbbm{1}+ H_{0}) ~\delta_{n m}\\
 H^{F}_{1,nm} &= H_1~ (\delta_{n m+1} + \delta_{n m-1})\\
 H^{F}_{2,nm} &= H_2 ~(e^{i \varphi} \delta_{n m+1} + e^{-i \varphi} \delta_{n m-1})
\end{align}
and
\begin{align}
H_0 &= \begin{pmatrix}
      V_s & -2 (J_{s,o}+ J_{s,e}  e^{-2 i k}) \\
      -2 (J_{s,o}+ J_{s,e} e^{2 i k})& -V_s 
      \end{pmatrix} \\
H_1 &= \begin{pmatrix}
      0 & - (J_o+ J_e  e^{-2 i k}) \\
      - (J_o + J_e e^{2 i k}) & 0
      \end{pmatrix} \\
H_2 &= \begin{pmatrix}
      \frac{V_o}{2} & 0 \\
      0 & -\frac{V_e}{2}
      \end{pmatrix}   
\end{align}
Similarly, the doublon velocity can be expressed in terms of the perturbation
\begin{equation}
 v_d^{F} = v^{F}_0 + \epsilon_1 v^{F}_1
\end{equation}
with
\begin{align}
 v^{F}_{0,nm} &= v_{0} \delta_{n m}\\
 v^{F}_{1,nm} &= v_{1} (\delta_{n m+1} - \delta_{n m-1})
\end{align}
and
\begin{align}
v_0 &= i \begin{pmatrix}
      0 & -2 (J_{s,o} - J_{s,e}  e^{-2 i k})\\
      2 (J_{s,o}- J_{s,e}  e^{2 i k}) & 0
      \end{pmatrix} \\
v_1 &= i \begin{pmatrix}
      0 & - (J_o- J_e  e^{-2 i k})\\
      (J_o- J_e  e^{2 i k}) & 0
      \end{pmatrix}
\end{align}
The Floquet-type Schr\"odinger equation reads
\begin{equation}
 (H^{F}_0 + \epsilon_1 H^{F}_1 + \epsilon_2 H^{F}_2) |n \alpha \rangle  = E_{n\alpha}|n \alpha \rangle
\end{equation}
where $(n\alpha)$ indicates the basis that diagonalises the Floquet Hamiltonian $H^{F}$, with $n$ describing the Floquet mode and $\alpha$ being the index of the original Hilbert space. 

\begin{figure}[t!]
    \centering
    \includegraphics[width=8cm]{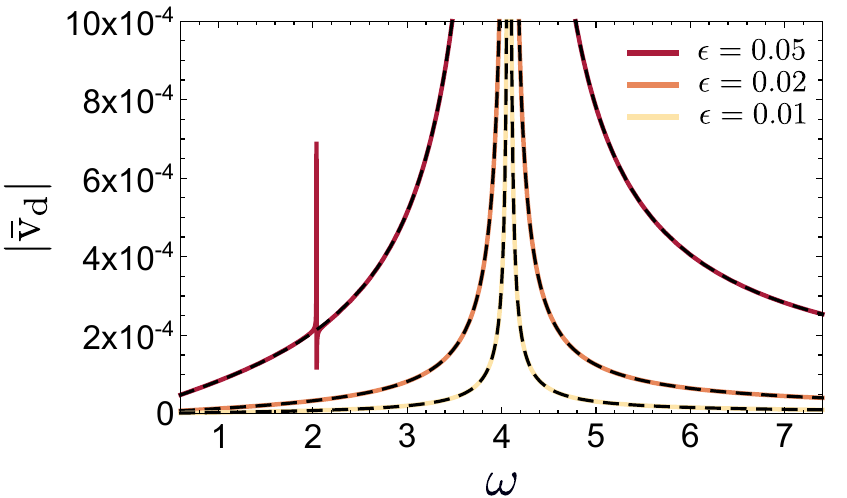}
    \caption{Absolute values of doublon velocities for the main resonance for different values of $\epsilon=\epsilon_1=\epsilon_2$ (other parameters as in Fig.~\ref{fig:fig2}). 
    Coloured lines represent exact numerical data and corresponding dashed lines are plots of (\ref{eq:pert_result}).
    Sharp spike appearing on curve for $\epsilon=0.05$ corresponds to second resonance, which is not captured by (\ref{eq:pert_result}).}
    \label{fig:resonance}
\end{figure}

As we are interested in the particle velocity $\langle v_d^{F} \rangle$, corrections to the eigenstates are of interest only.
Equating components of same power in $\epsilon_1,\epsilon_2$ leads to following corrections
\begin{align}
 |n \alpha^{1} \rangle &= \sum_{\substack{m\beta\neq \\ n\alpha}} \frac{\langle m \beta | H_{1}^{F} | n\alpha \rangle} {E_{n\alpha} - E_{m\beta}}| m\beta \rangle \label{eq:na1} \\
 |n \alpha^{2} \rangle &= \sum_{\substack{m\beta\neq \\ n\alpha}}  \frac{\langle m \beta | H_{2}^{F} | n\alpha \rangle} {E_{n\alpha} - E_{m\beta}}| m\beta \rangle  \label{eq:na2}\\
 |n \alpha^{12} \rangle &= \sum_{\substack{l\gamma\neq \\ n\alpha}}  \sum_{\substack{m\beta\neq \\ n\alpha}} \frac{\langle l \gamma | H_{1}^{F} | m\beta \rangle \langle m \beta | H_{2}^{F} | n\alpha \rangle}
 {(E_{n\alpha} - E_{l\gamma})(E_{n\alpha} - E_{m\beta})}| m\beta \rangle  \label{eq:na12} \\ \nonumber
  &\quad + \frac{\langle l \gamma | H_{2}^{F} | m\beta \rangle \langle m \beta | H_{1}^{F} | n\alpha \rangle}
 {(E_{n\alpha} - E_{l\gamma})(E_{n\alpha} - E_{m\beta})}| m\beta \rangle
 \end{align}
Note that all energies and eigenstates appearing in (\ref{eq:na1})-(\ref{eq:na12}) are zero order expansion coefficients (i.e., $E_{n\alpha}=E_{n\alpha}^{0}$ etc.).
We want to calculate $\langle v_d^{F} \rangle$ for a state with zero inital momentum, $k=0$.
In this case, all first order terms vanish by construction (together with the zeroth order term).
We, thus, want to constrain ourself to second order perturbation theory. 
Notice further that due to symmetry reasons only $\epsilon_1\epsilon_2$-terms yield a finite contribution.
Hence, the velocity can be approximated by
\begin{equation}
\begin{split}
\label{eq:currentexp}
 \langle n\alpha | v_d^{F} | n \alpha \rangle \approx \epsilon_1\epsilon_2 \big( \langle n\alpha^{0} | v_0 | n \alpha^{12} \rangle +  \langle n\alpha^{1} | v_0 | n \alpha^{2} \rangle \\
   + ~ \langle n\alpha^{0} | v_1 | n \alpha^{2}  \rangle+ h.c. \big)
 \end{split}
\end{equation}
Note that here we determine the velocity in the adiabatic case.
Plugging Eqs.~(\ref{eq:na1})-(\ref{eq:na12}) into (\ref{eq:currentexp}) generates many terms that can be put in the following compact form 
for our initial $2\times2$ problem
\begin{equation}
\label{eq:pert_result}
 \begin{split}
  \langle v_d^{F} \rangle  &\approx \epsilon_1 \epsilon_2 \sin(\varphi)  \times \bigg[ 2 v_{1,\gamma\alpha} H_{2,\gamma\alpha} \left(
   \frac{1}{ \delta -\omega} - \frac{1}{\omega + \delta}\right) \\
   &+  \frac{8 v_{0,\gamma\alpha} \omega }{\delta \omega^{2}-\delta^{3}} 
   \left(H_{1,\gamma\alpha}H_{2,\alpha\alpha}-H_{2,\gamma\alpha}H_{1,\alpha\alpha}\right) \bigg]
   \end{split}
\end{equation}
with $\delta= \varepsilon_\gamma - \varepsilon_\alpha$, $H_{1,\gamma\alpha} = \langle \gamma |H_{1}| \alpha \rangle$ etc. and $\alpha,\gamma \in \{1,2\}$ (1: lower band, 2: upper band), 
but $\gamma \neq \alpha$.
The correct expression for the velocity of holes is gained by inserting the modified versions of $H^F$ and $v^F$.
Clearly, expression (\ref{eq:pert_result}) breaks down at the resonance, where it predicts a pole for the velocity.
Therefore, only the ``wings'' are compared to exact numerical data in Fig.~\ref{fig:resonance}.
We compare absolute values of the velocity, 
because perturbation theory will yield the opposite band when sweeping through the resonance (c.f. Fig.~\ref{fig:avcross}) and hence will give an undesired sign change in the velocity.

In the limit $\omega \gg \varepsilon_{\alpha/\gamma}$ Eq.~(\ref{eq:pert_result}) can be used to describe the velocity for high frequencies even for $\epsilon_1=\epsilon_2=1$. 
The expression simplifies in this case to 
\begin{equation}
\label{eq:pert_largew}
 \begin{split}
   \langle n\alpha | v_d^{F} | n \alpha \rangle & \overset{\omega \gg \varepsilon_{\alpha/\gamma}}{\approx} 
   \frac{\epsilon_1 \epsilon_2 \sin(\varphi)}{\omega}  \times \bigg[ 4 v_{1,\alpha\gamma} H_{2,\gamma\alpha} \\
   & +  \frac{8 v_{0,\gamma\alpha}}{\delta} \left(H_{1,\gamma\alpha}H_{2,\alpha\alpha}-H_{2,\gamma\alpha}H_{1,\alpha\alpha}\right) \bigg]
   \end{split}
\end{equation}
Due to the property that $H_2 \to - H_2$ when switching from doublons to holes, one immediately sees that the velocities (and hence the effective vector potentials) of the two species must have opposite signs in the large frequency limit.

\section{Degenerate perturbation theory}
\label{appDPT}
In order to be able to describe the behaviour around the resonance correctly, one needs to apply \textit{degenerate} perturbation theory.
Degeneracies leading to the main resonance of the system can be found if $\delta = \omega$, with  $\delta= \varepsilon_\gamma - \varepsilon_\alpha$.
Going into the basis that diagonalises $H_{0}^{F}$ one defines the degeneracy subspace
\begin{align}
\tilde{H}_0 &=& \begin{pmatrix}
     -\frac{1}{2}(\omega - \delta ) & (\epsilon_1 H_1 + \epsilon_2 e^{-i \varphi}  H_2)_{21} \\
     (\epsilon_1 H_1 + \epsilon_2 e^{i \varphi}  H_2)_{12}  & \frac{1}{2}(\omega - \delta )
      \end{pmatrix}  
\end{align}
Here, the notation $(.)_{\alpha\beta}$ refers to the respective operator in the basis $\{|n \alpha^{0}\rangle\}$ with subspace index $(\alpha,\beta)$.
The eigenvalues of the matrix are $\{-E(\omega),E(\omega)\}$, where $E(\omega) \ll \omega$ for $\omega\approx\varepsilon_{\gamma}-\varepsilon_{\alpha}$.  
Alternatively, the Hamiltonian $\tilde{H}_0$ can be written in the convenient form
\begin{equation}
\tilde{H}_0 = \bf\tilde{{h}}_{0} \cdot \boldsymbol{\sigma}
\end{equation}
where $\boldsymbol{\sigma}$ is the vector of Pauli matrices and $\tilde{\bf{h}}_{0} = ( \tilde{h}_{x},\tilde{h}_{y},\tilde{h}_{z} )$ with
\begin{align}
\label{hx}
 \tilde{h}_{x} &= \epsilon_1 H_1 + \epsilon_2 \cos(\varphi) (H_2)_{21} \\
 \label{hy}
 \tilde{h}_{y} &= \epsilon_2 \sin(\varphi) (H_2)_{21} \\
 \label{hz}
 \tilde{h}_{z} &= - \frac{1}{2}(\omega -\delta )
 \end{align}
Here we used that $(H_1)_{12}=(H_1)_{21}$ and $(H_2)_{12}=(H_2)_{21}$.

\begin{figure}[t!]
    \centering
    \includegraphics[width=8cm]{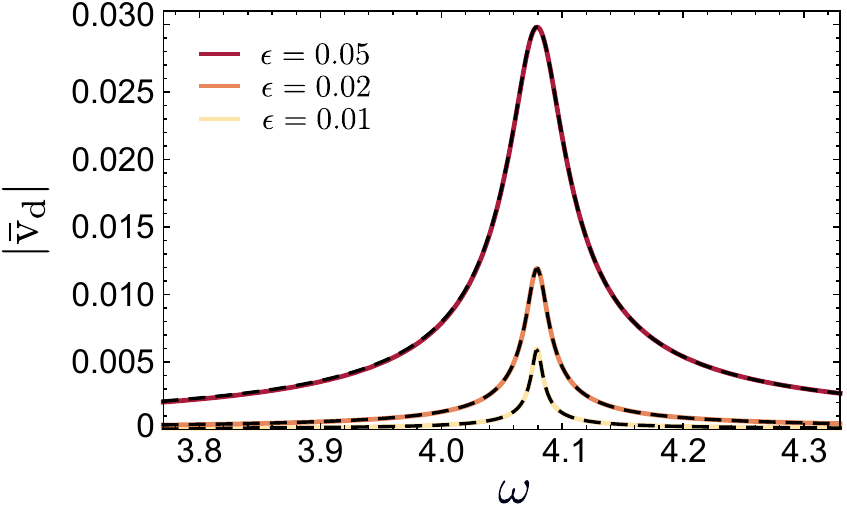}
    \caption{Doublon velocities around the main resonance for different values of $\epsilon=\epsilon_1=\epsilon_2$ (other parameters as in Fig.~\ref{fig:fig2}). 
    Coloured lines represent exact numerical data and corresponding dashed lines are plots of (\ref{eq:degpert_result}).
    }
    \label{fig:resonance2}
\end{figure}

Dividing the original Floquet Hamiltonian into a structure composed of such degeneracy blocks, allows to define effective coupling matrices.
In the following we shall only consider those which couple next-nearest degeneracy blocks, namely:
\begin{align}
\tilde{H}_1 &= \begin{pmatrix}
       (H_1)_{22} & 0\\
      0 & (H_1)_{11}
      \end{pmatrix} 
      = (H_1)_{22} \sigma_{z}  \\
\tilde{H}_2 &= e^{i \varphi} \begin{pmatrix}
      (H_2)_{22} & 0 \\
      0 & (H_2)_{11}
      \end{pmatrix} 
      = e^{i \varphi} (H_2)_{22} \sigma_{z}   
\end{align}
where $\tilde{H}_1,\tilde{H}_2$ $(\tilde{H}_{1}^{\dagger},\tilde{H}_{2}^{\dagger})$ describe the coupling to a higher (lower) Floquet mode.
Here we used the property that $(H_1)_{11}=- (H_1)_{22}$ and $(H_2)_{11}=- (H_2)_{22}$, and that $H_1$,$H_2$ are real matrices for $k=0$.
The same procedure is done for the respective parts of the (doublon) velocity operator
\begin{align}
\tilde{v}_1&= \begin{pmatrix}
      0 &  (v_1)_{21} \\
       (v_1)_{12} & 0
      \end{pmatrix} 
      = i (v_1)_{21} \sigma_{y}\\
\tilde{v}_0 &= \begin{pmatrix}
      0 & (v_{0})_{21}\\
      0 & 0
      \end{pmatrix}
      =  \frac{(v_1)_{21}}{2} (\sigma_x + i \sigma_{y})
\end{align}
where $(v_1)_{21} = -(v_1)_{12}$.
In order to find the velocity one first determines the basis states of $\tilde{H}$, i.e., $|\tilde{n}\tilde{\alpha} \rangle$, 
by performing again standard perturbation theory as in the previous section (see Eqs.~(\ref{eq:na1})-(\ref{eq:na12})).
The velocity in this picture is approximately given by
\begin{equation}
\label{vF1}
\begin{split}
\langle \tilde{n}\tilde{\alpha} |v^{F}_d| \tilde{n}\tilde{\alpha} \rangle \approx \epsilon_1 \langle \tilde{n}\tilde{\alpha}^{0} | \tilde{v}_1 | \tilde{n}\tilde{\alpha}^{0}\rangle +
 (\epsilon_1 \langle \tilde{n}\tilde{\alpha}^{0} | \tilde{v}_0 | \tilde{n}\tilde{\alpha}^{1}\rangle  \\
+ \epsilon_2 \langle \tilde{n}\tilde{\alpha}^{0} | \tilde{v}_0 | \tilde{n}\tilde{\alpha}^{2}\rangle +h.c.)
\end{split}
\end{equation}
To evaluate this expression we exploit the fact that $\langle \tilde{\alpha} | \sigma_{n} | \tilde{\alpha} \rangle = \pm \tilde{h}_{n}/|\bf{\tilde{h}_0}|$,
where the $+(-)$ indicates the upper (lower) band in the degenerate subspace 
(note that the upper band in this subspace corresponds to the lower band in the original Hamiltonian, hence we adjust our notations such that $\tilde{\alpha}$ can be identified with $\alpha$ in Eq.~(\ref{eq:pert_result}))
Finally, the evaluation of Eq.~(\ref{vF1}) yields
\begin{equation}
 \begin{split}
\langle v^{F}_d &  \rangle  \approx 
\pm \frac{1}{|\bf{\tilde{h}_0}|} \Big[- \epsilon_1 (v_1)_{21} \tilde{h}_y+ \frac{2}{\omega} \epsilon_1 (v_0)_{21} (H_1)_{22} \tilde{h}_y   \\
&+ \frac{2}{\omega} \epsilon_2 (v_0)_{21} (H_2)_{22} \left( \cos(\varphi) \tilde{h}_y - \sin(\varphi) \tilde{h}_x  \right) \Big]
 \end{split}
\end{equation}
where the $+(-)$ refers now to the lower (upper) band of the original Floquet Hamiltonian, $H^{F}$.
Using relations (\ref{hx})-(\ref{hz}) the expression for the velocity simplifies further to read
\begin{equation}
\label{eq:degpert_result}
 \begin{split}
  &\langle v^{F}_d \rangle \approx \pm ~ 2~ \epsilon_1 \epsilon_2 \sin(\varphi) \times \bigg[ \frac{H_{2,21}v_{1,12}}{\sqrt{ (\delta -\omega)^2 +\Delta^2 }} \\
&+ \frac{2 ~ v_{0,21}}{\omega \sqrt{ (\delta -\omega)^2 +\Delta^2 }} (H_{1,22} H_{2,21}- H_{1,21} H_{2,22})\bigg]
  \end{split}
\end{equation}
with
\begin{equation}
 \Delta^2 = 4 \left( \epsilon_1^2 H_{1,21}^2 + \epsilon_2^2 H_{2,21}^2 + 2 \epsilon_1 \epsilon_2 \cos(\varphi) H_{1,21} H_{2,21}  \right)
\end{equation}
Comparing Eqs.~(\ref{eq:pert_result}) and (\ref{eq:degpert_result}), one sees that the major modification is the appearance of the $\Delta$-term, turning the pure pole into a Lorentzian-like function.
In Fig.~\ref{fig:resonance2} exact numerical velocities are shown together with plots of Eq.~(\ref{eq:degpert_result}) for different values of $\epsilon$.

In order to investigate the height of the velocity peak one sets $\omega=\delta$.
For the special case that $\epsilon_1=\epsilon_2=\epsilon$ one observes that $\langle v^{F}_d \rangle \approx \mathcal O(\epsilon^2)/\mathcal O(\epsilon)=\mathcal O(\epsilon)$.
Hence, the height of the main resonance is directly proportional to $\epsilon$ (see Fig. \ref{fig:resonance2}).
Generally, the height of the $n^\text{th}$ resonance is given by $\mathcal O(\epsilon^n)$.
As before, the analysis of the hole velocity is completely analogous.

\section{Explicit calculation of the effective vector potential}
\label{appEVP}

\begin{figure}[t!]
    \centering
    \includegraphics[width=8cm]{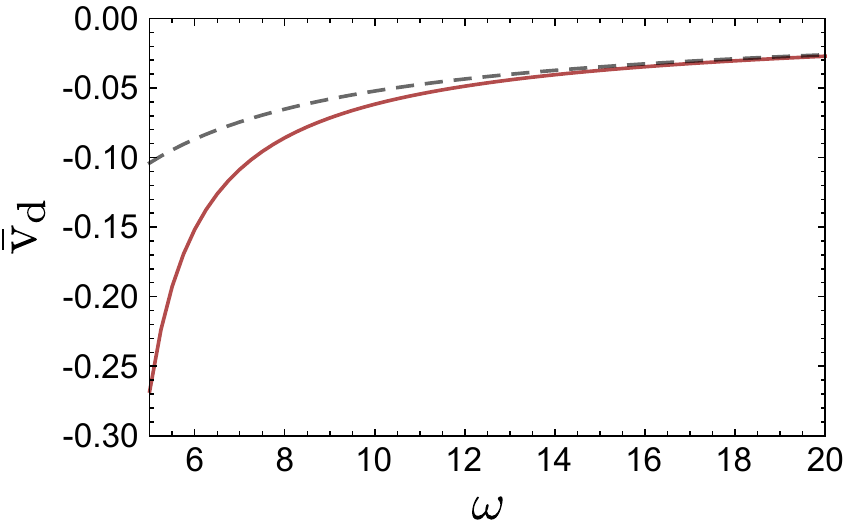}
    \caption{Comparison of doublon velocities of exact numerical data (red/solid) and plots of Eq.~\ref{veleff} (grey/dashed) for large frequencies $\omega > J_o,J_e,V_o,V_e$ 
    (parameters as in Fig.~\ref{fig:fig2}).
    }
    \label{fig:veleff}
\end{figure}

In the high frequency regime, $\omega\gg J_o,J_e,V_o, V_e$, the time-evolution operator for one period of driving, $U(T)$, can be approximated by a truncated Magnus expansion
\begin{equation}
 \begin{split}
U(T) \approx  \exp&\Big( - i \int_0^T dt_1 H_d(t_1) \\
&- \frac{1}{2} \int_0^T dt_1 \int_0^{t_1} dt_2 [H_d(t_1),H_d(t_2)]\Big)   
 \end{split}
\end{equation}
Using Eq.~\ref{momentumB} and performing the time integrals explicitly, 
one can (similar to Ref.~\cite{Grushin2014}) write the effective static (doublon) Hamiltonian (\ref{heffF}) approximated to first order in $1/\omega$ as
\begin{equation}
 H^{\rm eff}_{d} = H^0_{d} - \frac{1}{\omega} ( [H^0_{d},H^{-1}_{d}] - [H^0_{d},H^{1}_{d}] + [H^{-1}_{d},H^{1}_{d}] )
 \label{Heff}
\end{equation}

After inserting expressions (\ref{H0}) and (\ref{H1}) and appropriate simplification, the effective Hamiltonian can be written as
\begin{equation}
 H^{\rm eff}_{d} =
 \begin{pmatrix}
  V_{s} & C \\
  C^* & - V_{s}
 \end{pmatrix}
\end{equation}
with
\begin{align}
\begin{split}
C &= - 2 (J_{s,o} + J_{s,e} e^{-2ik}) - \frac{i \sin(\varphi)}{\omega} (V_{o}-V_{e}) \\  
&\phantom{=- 2J_{s,o}}\times(-2J_{s,o}+J_o - (2 J_{s,e}- J_e) e^{-2 i k})
\end{split}\\
&\approx -2 (J_{s,o} e^{-i a} + J_{s,e} e^{-ib} e^{-2ik})
\end{align}
where
\begin{align}
 a = \frac{1}{\omega} \sin(\varphi) (V_{o} - V_{e}) \left(1 + \frac{J_{o}}{2 J_{s,o}}\right) \\
 b = \frac{1}{\omega} \sin(\varphi) (V_{o} - V_{e}) \left(1 + \frac{J_{e}}{2 J_{s,e}}\right)
\end{align}
Upon diagonalization of $H_d^{\rm eff}$, one can easily verify that the corresponding eigenenergies, $\varepsilon_{\alpha}^{\rm eff}$, are given as
\begin{equation}
\varepsilon_{\alpha}^{\rm eff}(k) = \varepsilon_{\alpha}^0(k-\mathcal A_d) 
\label{energyeff}
\end{equation}
with
\begin{equation}
\mathcal A_d = a-b = \frac{1}{2 \omega} \sin(\varphi) (V_o - V_e) \left( \frac{J_o}{J_{s,o}} - \frac{J_e}{J_{s,e}} \right)       
\end{equation}
being an effective vector potential coupling to the momentum $k$.
Similar expressions also hold for holes with modified parameters $V\to-V$ and $2J\to J$.

The resulting average (doublon) velocity associated with quasienergy band $\alpha$ is then conveniently given by the standard form
\begin{equation}
\bar{v}_{d,\alpha}^{\rm eff}(k) =  \frac{\partial \varepsilon_{\alpha}^0(k-\mathcal A_d)}{\partial k}
\label{veleff}
\end{equation}
In Fig.~\ref{fig:veleff} the velocity (\ref{veleff}) is compared to exact numerical data for $k=0$.
It is clearly visible how Eq.~(\ref{veleff}), and hence Eqs.~(\ref{Heff}) and (\ref{energyeff}), becomes valid in the large frequency regime.

\bibliographystyle{aipnum4-1}
\bibliography{bib}

\end{document}